\documentclass[iop]{emulateapj}
\shorttitle{Globular Clusters in NGC 4649}
\shortauthors{Strader et al.}
\def\etal{{\it et al.}}

\def\gsim{\;\rlap{\lower 2.5pt
 \hbox{$\sim$}}\raise 1.5pt\hbox{$>$}\;}
\def\lsim{\;\rlap{\lower 2.5pt
\hbox{$\sim$}}\raise 1.5pt\hbox{$<$}\;}
\usepackage{subfigure}
\slugcomment{To Appear in ApJ}

\begin{document}

\title{Deep Chandra Monitoring Observations of NGC 4649: II. Wide-Field \emph{Hubble Space Telescope} Imaging of the Globular Clusters}

\author{Jay Strader\altaffilmark{1,2}, Giuseppina Fabbiano\altaffilmark{2}, Bin Luo\altaffilmark{2}, Dong-Woo Kim\altaffilmark{2}, Jean P.~Brodie\altaffilmark{3}, Tassos Fragos\altaffilmark{2}, John S.~Gallagher\altaffilmark{4}, Vassiliki Kalogera\altaffilmark{5}, Andrew King\altaffilmark{6}, Andreas Zezas\altaffilmark{7}}

\email{strader@pa.msu.edu}

\altaffiltext{1}{Department of Physics and Astronomy, Michigan State University, East Lansing, Michigan 48824}
\altaffiltext{2}{Harvard-Smithsonian Center for Astrophysics, Cambridge, MA 02138}
\altaffiltext{3}{UCO/Lick Observatory, 1156 High St., Santa Cruz, CA 95064}
\altaffiltext{4}{Department of Astronomy, University of Wisconsin, Madison, WI 53706-1582}
\altaffiltext{5}{Center for Interdisciplinary Exploration and Research in Astrophysics (CIERA) \& Department of Physics and Astronomy, Northwestern University, 2145 Sheridan Road, Evanston, IL 60208}
\altaffiltext{6}{Department of Physics \& Astronomy, University of Leicester, University Road, Leicester LE1 7RH UK}
\altaffiltext{7}{Physics Department, University of Crete, P.O. Box 2208, GR-710 03, Heraklion, Crete, Greece}

\begin{abstract}

We present $g$ and $z$ photometry and size estimates for globular clusters (GCs) in the massive Virgo elliptical NGC 4649 (M60) using a five-pointing \emph{Hubble Space Telescope}/Advanced Camera for Surveys mosaic. The metal-poor GCs show a monotonic negative metallicity gradient of $-0.43\pm0.10$ dex per dex in radius over the full radial range of the data, out to $\sim 24$ kpc. There is evidence for substantial color substructure among the metal-rich GCs. The metal-poor GCs have typical sizes $\sim 0.4$ pc larger than the metal-rich GCs out to large galactocentric distances ($\ga 20$ kpc), favoring an intrinsic explanation for the size difference rather than projection effects. There is no clear relation between half-light radius and galactocentric distance beyond $\sim 15$ kpc, suggesting that the sizes of GCs are not generically set by tidal limitation. Finally, we identify $\sim 20$ candidate ultra-compact dwarfs that extend down to surprisingly faint absolute magnitudes ($M_z \sim -8.5$), and may bridge the gap between this class and ``extended clusters" in the Local Group. Three of the brighter candidates have published radial velocities and can be confirmed as bona fide ultra-compact dwarfs; follow-up spectroscopy will determine the nature of the remainder of the candidates.

\end{abstract}


\section{Introduction}

It is well-established that a large fraction of low-mass X-ray binaries in early-type galaxies are dynamically formed in globular clusters (GCs; see the review of Fabbiano 2006). These sources are preferentially found in red metal-rich GCs and in those clusters with the highest interaction rates, which occur in the clusters with small radii and large stellar densities (e.g., Kundu \etal~2002; Pooley \etal~2003). Therefore, precision photometry and structural parameters for GCs are necessary to properly interpret X-ray observations of nearby elliptical galaxies.

To complement new deep, multi-epoch Chandra X-ray observations of the massive Virgo elliptical NGC 4649 (M60; VCC 1978), we have obtained \emph{Hubble Space Telescope}/Advanced Camera for Surveys ($HST$/ACS) pointings that tile most of the Chandra field of view. The Chandra catalog is presented in Luo \etal~(2012). Future papers will discuss the properties of low-mass X-ray binaries in NGC 4649, specific interesting systems, and other topics.

This paper derives properties of GCs (photometry and sizes) in the $HST$/ACS data and points out several interesting features of the GC system itself, in the context of their use as tracers of the formation and evolution of massive galaxies (Brodie \& Strader 2006). The most notable feature of this dataset, compared to the voluminous set of previous $HST$ imaging studies of massive ellipticals, is the areal coverage: our five ACS pointings, combined with an archival pointing of equivalent depth, cover $\sim 60$ square arcmin, with near-complete azimuthal coverage to projected radii $\ga 20$ kpc. These data allow the study of GC colors and effective radii as a function of galactocentric radius over a radial range not usually possible with $HST$ data, unless one studies more distant galaxies. Among the scientific questions that can be addressed is whether the well-known size difference between blue and red GCs is due primarily to projection effects (e.g., Larsen \& Brodie 2003), and if metallicity gradients are present within the individual GC subpopulations (Harris 2009; Forbes \etal~2011). The added areal coverage also aids in the hunt for uncommon objects, such as candidate ultra-compact dwarf galaxies (Brodie \etal~2011).

\section{Data Reduction and Analysis}

\subsection{Basic Reductions}

The ACS observations were arranged in a five-pointing mosaic with slight overlap around a central archival pointing from the ACS Virgo Cluster Survey (Cote \etal~2004). Together, these data cover the sensitive portion of the Chandra ACIS field of view, excluding the nearby spiral NGC 4647. Each pointing consists of 888 sec of total exposure time in F475W and 1278 sec in F850LP (henceforth called $g$ and $z$), with each exposure split using a simple line dither to cover the ACS chip gap. The spatial coverage of the data is shown in Figure 1.

\begin{figure}
	\epsscale{1.23}
	\plotone{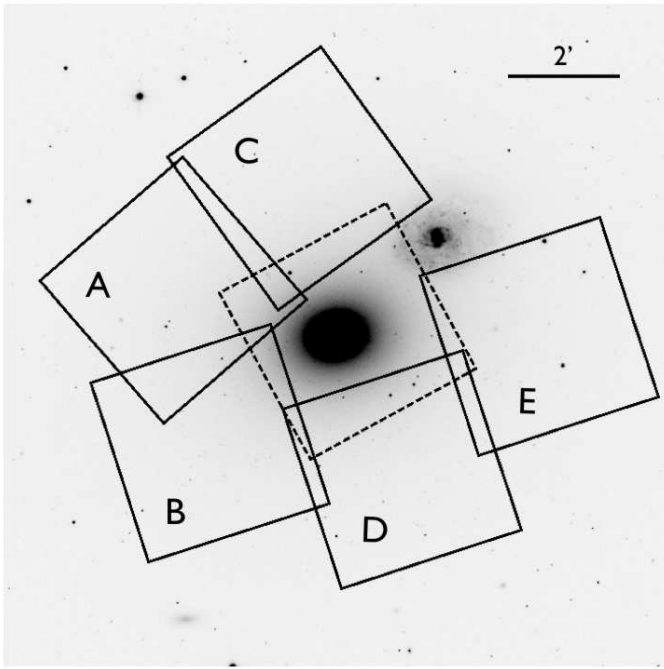}
\figcaption[z_good.eps]{\label{fig:fig_1}
An SDSS DR7 mosaic image of NGC 4649 (and the foreground spiral NGC 4647), with the approximate location of our ACS pointings superposed (solid lines). The archival ACS Virgo Cluster Survey central pointing is shown with dotted lines.}
\end{figure}

ACS observations taken after the most recent $HST$ servicing mission require additional processing steps that, at the time of these reductions, were not yet incorporated into the standard pipeline. We started with the {\tt .flt} files that have been bias and dark-corrected and flat-fielded. First, we corrected the bias striping using the PyRaf routine provided by STScI.\footnote{http://www.stsci.edu/hst/acs/software/destripe} Next, we corrected for the ACS charge transfer efficiency losses using the STScI routine that implements the algorithm of Anderson \& Bedin (2010).\footnote{http://www.stsci.edu/hst/acs/software/CTE/}. To  reject cosmic rays, we ran L.~A.~Cosmic (van Dokkum 2001) on the resulting images using relatively conservative criteria. Finally, these processed images were combined using MultiDrizzle in PyRaf to produce distortion-corrected mosaics for photometry for each filter and pointing.

Before performing photometry, we used moderately bright, unsaturated objects to shift the astrometry of each image onto the standard system of the Sloan Digital Sky Survey Data Release 7 (Abazajian \etal~2009).

\subsection{Cluster Detection and Photometry}

We used SExtractor (Bertin \& Arnouts 1996) to produce initial candidate lists for photometry. Sources were selected on median-subtracted images, using an optimal weight map, with a low threshold ($1.5\sigma$) and a requirement of at least five connected pixels. Separate candidate lists for the $g$ and $z$ images were matched to within 2 pixels (0.1\arcsec). 

We performed aperture photometry on this matched candidate list using a 5-pixel aperture. GCs at the distance of NGC 4649 are marginally resolved and larger clusters will have a greater fraction of their total flux distributed beyond a fixed aperture. We corrected the 5-pixel base magnitudes to 10-pixel (0.5\arcsec) magnitudes using size-dependent aperture corrections, calculated as described below in \S 2.3.1. These magnitudes were then corrected to a nominal infinite aperture using the values from Sirianni \etal~(2005). For the largest objects, an additional correction for light beyond the (0.5\arcsec) aperture was also required. Finally, these magnitudes were corrected for foreground reddening using the maps of Schlegel \etal~(1998) as updated by Peek \& Graves (2008). All magnitudes are on the AB system.

We have 234 GCs in common with the catalog of Jord{\'a}n \etal~(2009; these data were also discussed in Peng \etal~2006 and Mieske \etal~2006). The median difference in both $g$ and $z$ magnitudes between this catalog and our measurements is 0.01 mag. The median difference in $g-z$ is 0.002 mag. These comparisons show that our photometry is in excellent agreement with the published values, and all of the measurements can be compiled into a self-consistent photometric catalog. This catalog is given in Table 1. We assume a distance of 16.5 Mpc to NGC 4649 (Blakeslee \etal~2009) for the paper.

\subsection{Cluster Sizes}

We measured half-light radii using \emph{ishape} (Larsen 1999). Since cluster concentration $c$ cannot be reliably measured except for the most luminous clusters\footnote{$c = r_t/r_0$ for tidal radius $r_t$ and King radius $r_0$.}, we used a fixed value of $c = 30$ for all objects. For most clusters we used a fitting radius of 5 pixels (equivalent to 20 pc). For the largest clusters ($r_h > 10$ pc), we increased the fitting radius in 1-pixel increments as necessary so that it was always at least twice as large as the measured effective radius. 

Harris (2009) argues that measurements of $r_h$ can only be made with confidence for objects with $S/N >$ 50. For clusters of typical size and background, this value corresponds to magnitude limits of $z < 23$ and $g < 24$; we restrict our size measurements to these limits. In Table 1 we list the individual $g$ and $z$  $r_h$ values as well as a weighted average of the two bands.

There are 703 objects with size measurements in both $g$ and $z$. The median difference in $r_h$, in the sense of $r_{h,g} - r_{h,z}$, is 0.11 pc, indicating good agreement between the filters. Figure 2 shows a comparison between these size estimates. The random uncertainties in the sizes, given the assumed model (King profiles with fixed $c=30$)  were estimated by comparing clusters that fell within the boundaries of more than one pointing. Assuming the random uncertainties in $r_h$ in each pointing are Gaussian and identical, the distribution of pairwise $r_h$ differences, considered separately for each filter, should have a variance that is twice that of the individual measurements. For each filter, we estimated this quantity in magnitude bins and then fit a smooth exponential function to the data. These fits were used to assign uncertainties for each $r_h$ estimate.

\begin{figure}
	\epsscale{1.2}
	\plotone{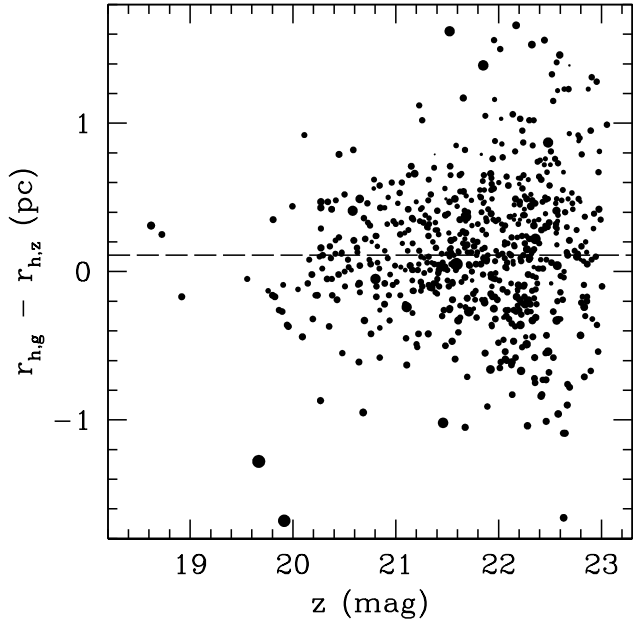}
\figcaption[sss.eps]{\label{fig:fig_2}
The differences between our $r_h$ measurements in $g$ and $z$ as a function of $z$ magnitude, with the point size proportional to the logarithm of $r_h$ in $z$. The median difference is marked with a dashed line.}
\end{figure}

For an individual filter, we find typical uncertainties of 10\% for more luminous objects and 20\% for the faintest clusters for which we report sizes. Previous studies using ACS data suggest that the systematic uncertainties in size measurements from point spread function (PSF) modeling for GCs at the distance of NGC 4649 are $\sim 0.4$ pc, or $\sim 20$\% for a typical cluster (Spitler \etal~2006; Harris 2009). The $r_h$ values in Table 1 include only the random measurement errors.

As with the photometry, we can compare our estimates of $r_h$ to those of Jord{\'a}n \etal~(2009), after converting their published $r_h$ values from angular to physical units using the assumed distance of 16.5 Mpc. In the median, our values are smaller by $0.25\pm0.04$ pc and $0.17\pm0.07$ in $g$ and $z$ respectively. These amount to systematic differences of $\sim 10$\%, well within the overall systematic uncertainty of size estimates from PSF modeling. In particular, Jord{\'a}n \etal~(2009) state systematic uncertainties in $r_h$ of $\sim 0.005$\arcsec (0.4 pc). The median differences in $r_h$ in both $g$ and $z$ are well within this estimate of the systematic uncertainty in the measurements. We compare $r_h$ measurements for GCs in common between this study and Jord{\'a}n \etal~(2009) in Figure 3. 

\begin{figure}
	\epsscale{1.2}
	\plotone{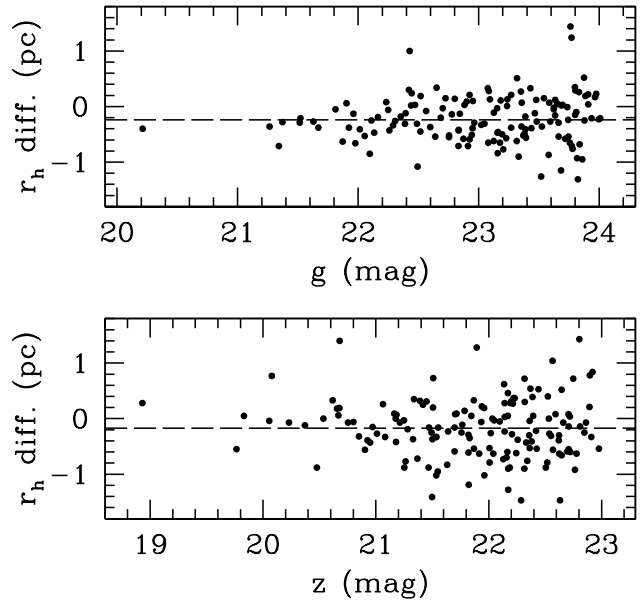}
\figcaption[fff7.eps]{\label{fig:fig_3}
The differences between our  $r_h$ estimates and those from Jord{\'a}n \etal~(2009), in the sense of us--them, as a function of $g$ and $z$ magnitude. The median difference is marked with a dashed line for each filter.}
\end{figure}

\subsubsection{Use in Aperture Corrections}

Our cluster $r_h$ estimates form the basis for size-dependent aperture corrections to the photometry. These were measured by convolving our empirical PSFs with King models of fixed concentration ($c=30$) and varying $r_h$. We define our basic aperture corrections from a radius of 5 pixels (0.25\arcsec) to 10 pixels (0.5\arcsec). For the largest clusters (with $r_h > 6$ pc), making up $\sim 3$\% of the total sample, an additional aperture correction for light falling outside of the 10 pixel radius is required. Representative corrections in $g$ are $-0.08$ and $-0.30$ mag for $r_h = 10$ and 20 pc, respectively. Because of the lower S/N of the empirical PSF in its outer regions, these extra corrections are less well constrained, and the total magnitudes of the largest clusters in our sample are less precise. However, these corrections have a minimal effect on the cluster colors (see below). 

Uncertainty in the value of $r_h$ for a particular cluster has a minimal effect on the $g-z$ color. For example, over the range 1--10 pc, the maximum deviation from the median correction is only 0.003 mag in $g-z$. For typical size uncertainties of 0.5--1 pc, the correction is utterly negligible. This point has been made before in the context of discussions of the ``blue tilt" of luminous metal-poor GCs (Jord{\'a}n \etal~2009). 

Uncertainties in sizes can have a larger effect on the total magnitudes of GCs. For clusters of typical size (2--3 pc), a 20\% size uncertainty translates into about 0.01 mag in luminosity. For larger clusters, the effect can be much more significant, corresponding to 0.04--0.05 mag for a cluster with $r_h = 10$ pc. However, few of the GCs are this large, and none of our conclusions depend strongly on the precise cluster luminosities in any case.

\subsection{Cluster Selection}

GCs were selected using these basic criteria: (i) $z < 24$; (ii) 0.5 $< g-z <$ 2.0. Any objects whose sizes were consistent with zero were presumed to be foreground stars and were discarded. No upper size limit was used because of the presence of candidate ultra-compact dwarf galaxies (see \S 3.3); in any case, few objects with sizes $> 10$ pc are present (22 in the whole catalog). The faint $z$ limit extends $\sim 1.3$ mag fainter than the turnover of the GC luminosity function at $z \sim 22.7$ (Jord{\'a}n \etal~2007). After these cuts, a visual inspection was used to remove obvious background galaxies.

Objects in the main body of the projected spiral NGC 4647 were also removed, though more distant GCs in this galaxy may remain in the catalog. We can roughly estimate this contamination as follows: a galaxy with $M_V \sim -19.9$ (de Vaucouleurs \etal~1991) should have $\sim 25-50$ GCs brighter than the turnover, about half of which are inside the excluded central $\sim 4-5$ kpc (Brodie \& Strader 2006). Further, our image mosaic only covers the half of the galaxy toward NGC 4649. Therefore, we expect that perhaps 5--10 objects in our catalog are GCs associated with NGC 4647.

The area covered is shown graphically in Figure 1, and is divided into our pointings (A--E) and the central ACS Virgo Cluster Survey pointing, denoted ``J" in our catalog.

About 8\% of our GCs were present in multiple pointings, and the photometry and size measurements for these objects are averages of the individual values. In addition, 20\% of the objects are present in the Jord{\'a}n \etal~(2009) catalog. For these objects, given the agreement between our photometric measurements, we simply average the magnitudes. For consistency, because of the small but significant differences in the size estimates, we do not average together the sizes, but solely use our estimates. We add the rest of the Jord{\'a}n \etal~(2009) candidates, applying the selection criteria above, into a master catalog with 1603 GCs. Of this total, 1120 have photometry from our new data.

\section{Results}

Figure 4 shows the $g-z$ vs.~$z$ color-magnitude diagram (CMD) of GCs in NGC 4649. The basic features of the the CMD are similar to those observed in previous $HST$/ACS studies of NGC 4649. Evident features include the standard bimodal color distribution in $g-z$ of blue metal-poor and red metal-rich GCs, and evidence for the ``blue tilt": a correlation between color and magnitude for the most luminous blue GCs (Strader \etal~2006; Harris \etal~2006; Mieske \etal~2006). We do not dwell on these established facts, but focus on the features of the GC system that can be studied with our improved radial coverage.

\begin{figure}
	\epsscale{1.2}
	\plotone{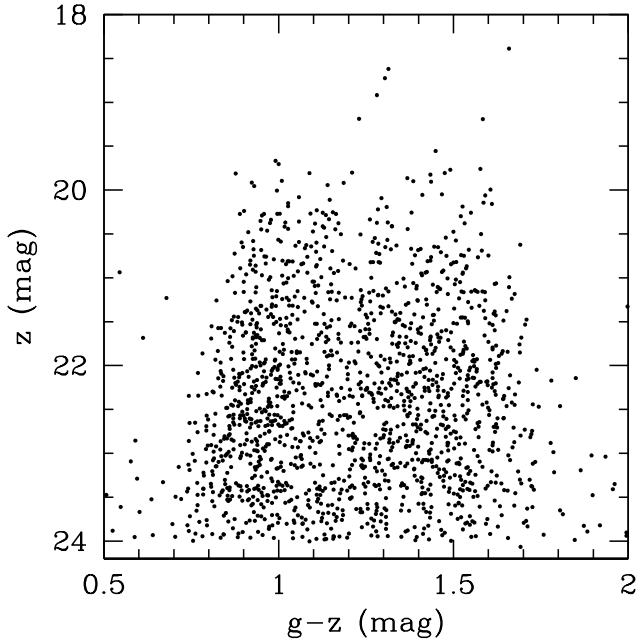}
\figcaption[n4649f1.eps]{\label{fig:fig_4}
$z$ vs.~$g-z$ color-magnitude diagram of GCs in NGC 4649. The standard blue and red GC subpopulations are visible, as well as a color--luminosity relation for the brighter blue (metal-poor) GCs.}
\end{figure}

\subsection{Radial Variations in Color}

The $g-z$ colors are plotted against the projected galactocentric distance in Figure 5. We overplot the peak colors of the blue and red subpopulations, calculated from density estimates using a Gaussian kernel of 0.05 mag and 1\arcmin\ (4.8 kpc) bins.\footnote{We use this method, rather than the common Gaussian mixture modeling, because it is less sensitive to outliers.} For these density estimates, only GCs with $23 < z < 21$ are used, to minimize the effects of the blue tilt and of faint contaminants.

\begin{figure}
	\epsscale{1.2}
	\plotone{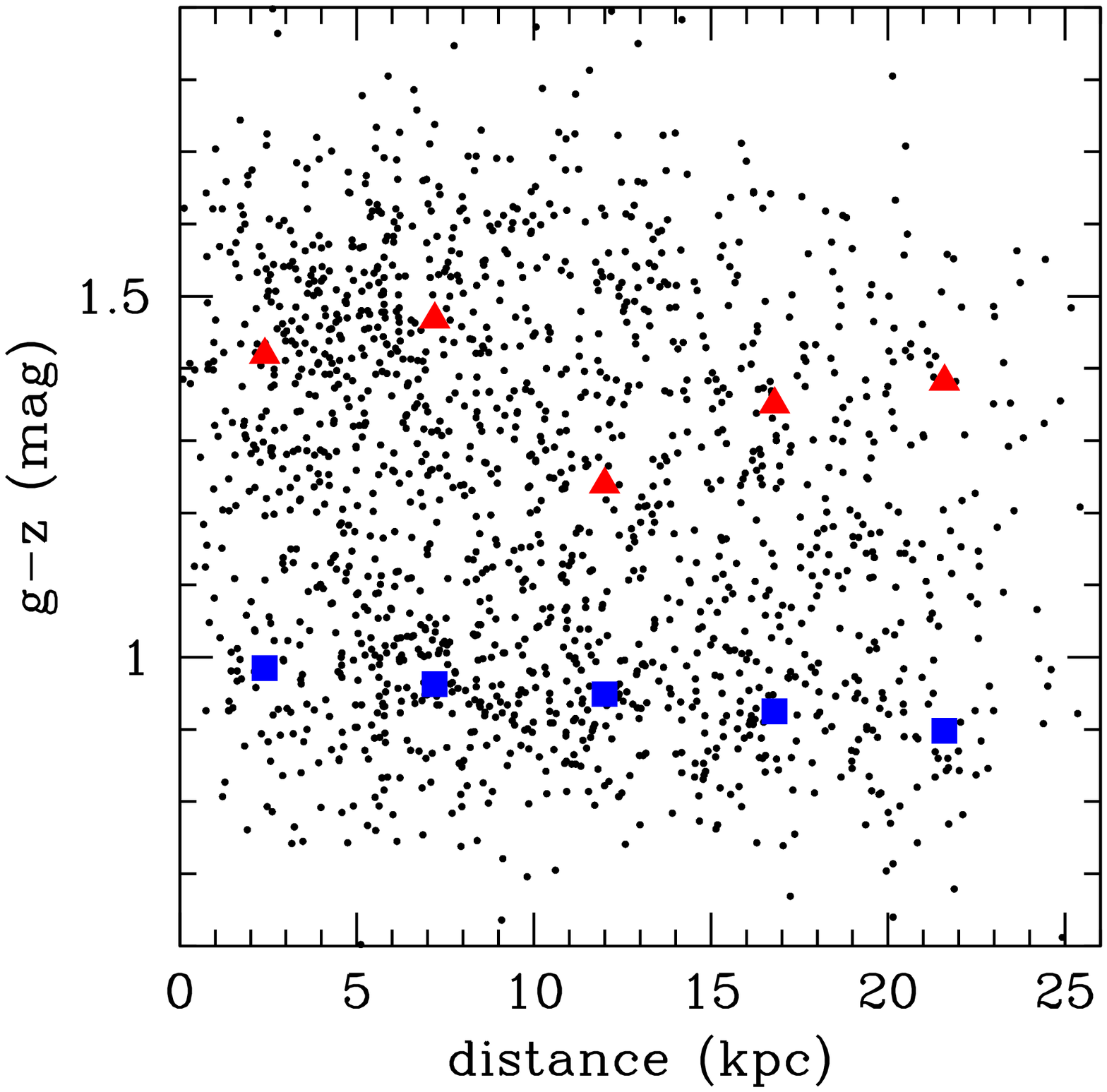}
\figcaption[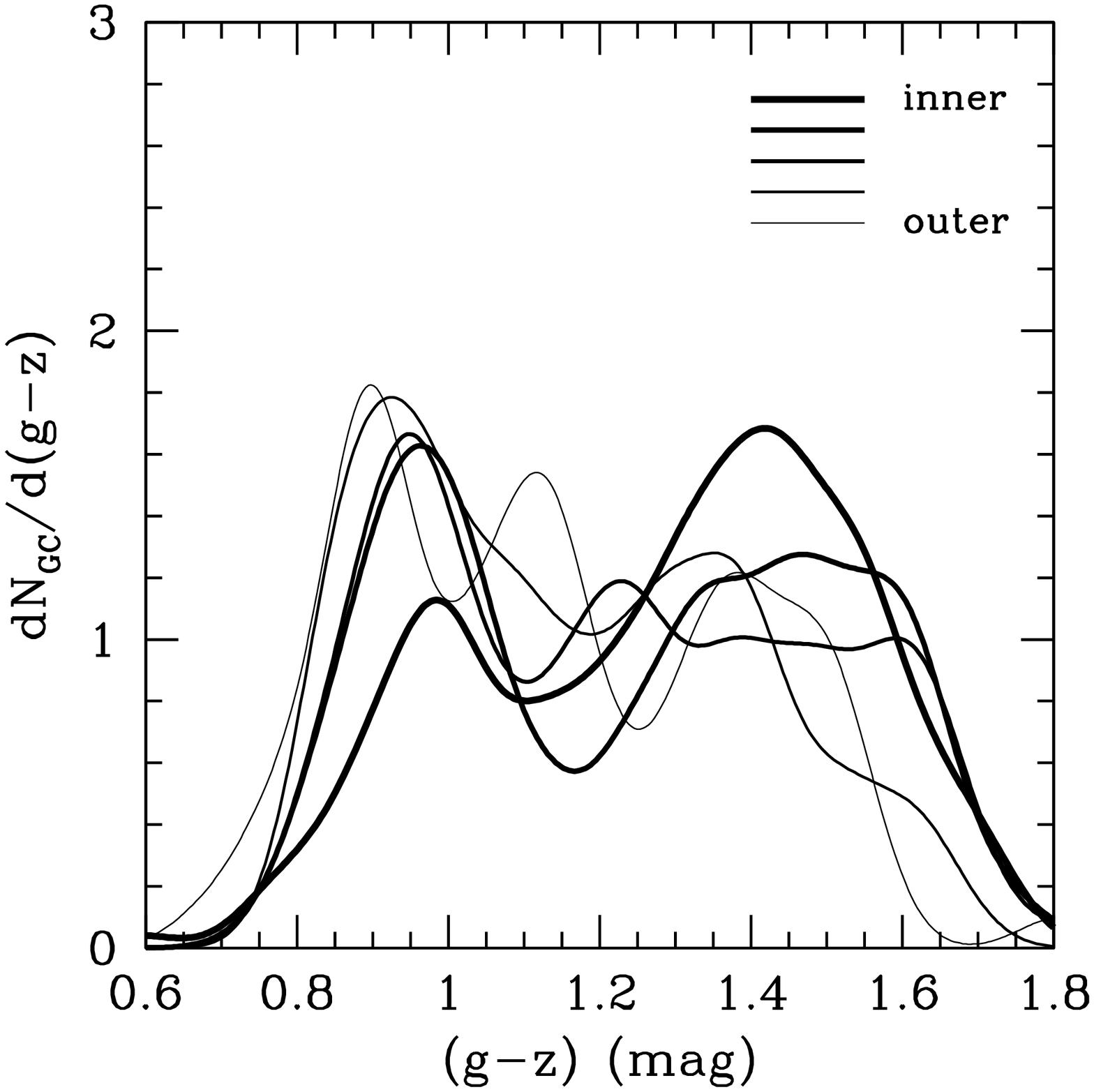]{\label{fig:fig_5}
$g-z$ vs.~projected galactocentric distance ($R_{\rm G}$) for GCs in NGC 4649. The overplotted blue squares and red triangles are the peak colors for the metal-poor and metal-rich subpopulations. There is a radial gradient in color (metallicity) for the metal-poor GCs, but no clear trend for the metal-rich GCs.}
\end{figure}

The metal-poor GCs have a monotonic color gradient with radius: $-0.021\pm0.001$ mag arcmin$^{-1}$, corresponding to a negative metallicity gradient of $-0.43\pm0.10$ dex per dex in radius (assuming the $g-z$ to [Fe/H] conversion of Peng \etal~2006). The outer edge of our data (at $\sim 24$ kpc) is close to the typical ``break radius" observed in the metal-poor GC systems of some other galaxies, beyond which the gradient flattens to near-zero (Harris 2001; Forbes \etal~2011), although this break appears to occur by $\sim 15$ kpc in M87 (Strader \etal~2011). Data at larger radii will be necessary to look for an edge to the gradient in NGC 4649.

We note that our measured gradient is somewhat steeper than found by Harris (2009) for metal-poor GCs in a sample of more distant brightest cluster galaxies. It not clear how comparable the two measurements are, as Harris necessarily restricted his fit to luminous GCs ($M_i < -9$) that are more likely to be affected by the blue tilt, which is known to vary with radius (Mieske \etal~2010). Indeed, if we restrict our analysis to GCs with $z<22$, yielding a luminosity cut similar to that of Harris (2009), there is no significant evidence for a metallicity gradient among the metal-poor GCs in NGC 4649. We conclude that the blue tilt can have an important effect on measured gradients and that conclusions based on the most luminous GCs may be biased.

There is minimal evidence for a metallicity gradient among the metal-rich GCs. While the mean peak color of the innermost two bins is $\sim 0.08$ mag redder in $g-z$ than the outermost two bins, there appears to be significant color substructure at all radii; at radii between 2\arcmin\ and 3\arcmin\ (9.6--14.4 kpc), there is no well-defined red peak at all. This substructure suggests the presence of GCs from accretion events that have not yet been fully phase mixed into NGC 4649.

Both the monotonic metal-poor GC gradient and the substructure in the red peak are evident in the density plots themselves, which are shown in Figure 6. These findings are remarkably similar to observations of GCs in the giant elliptical galaxies M87 (Strader \etal~2011; see their Figure 9) and NGC 1407 (Forbes \etal~2011; see their Figure 2). These data provide compelling evidence that the blue GC gradient and red GC substructure exist in the GC systems of many massive early-type galaxies.

The inner blue GC gradient appears to have a similar slope among the galaxies studied so far. The situation for the red GCs is less certain---while substructure in color appears ubiquitous, the presence of a red GC gradient is only clear in some galaxies, such as NGC 1407 (Forbes \etal~2011). In M87, Strader \etal~(2011) argue that there is an extended, kinematically-distinct intermediate-color subpopulation of GCs, separate from the standard red subpopulation, that causes the appearance of a gradient in the red GC color. These results suggest that care should be taken in the interpretation of simple two-population fits to GC color distributions.

\begin{figure}
	\epsscale{1.2}
	\plotone{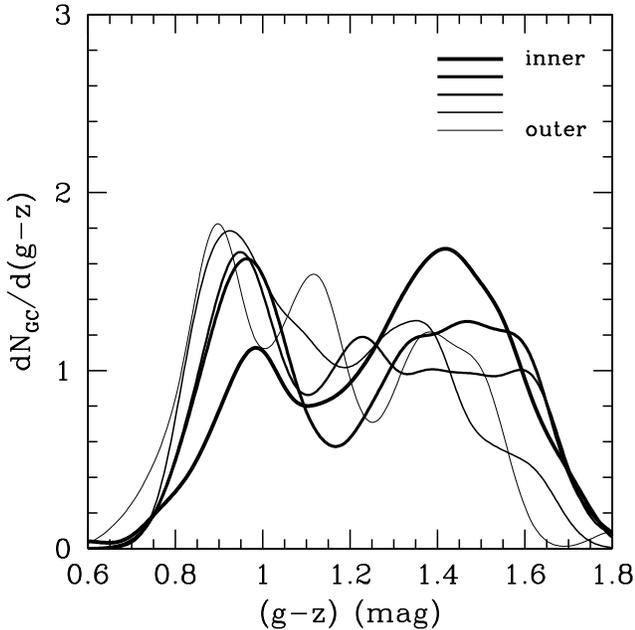}
\figcaption[f4.eps]{\label{fig:fig_6}
Kernel density estimates of the color distribution of GCs in NGC 4649 with $23 < z < 21$ as a function of projected galactocentric distance, with thicker lines representing more central distances. A fixed Gaussian kernel of 0.05 mag was used, and the mean distances plotted run from 0.5\arcmin\ (2.4 kpc) to 4.5\arcmin\ (21.6 kpc) in 1\arcmin\ (4.8 kpc) bins. The metal-poor peak is symmetric and shows the negative metallicity gradient discussed in the text, while the metal-rich peak shows significant color substructure at all distances.}
\end{figure}

\subsection{Radial Variations in Half-Light Radii}

Numerous previous studies of early-type galaxies, including NGC 4649, have shown that the metal-poor GCs are 0.3--0.5 pc larger than the metal-rich GCs (Kundu \& Whitmore 1998; Larsen \etal~2001; Jord{\'a}n \etal~2005), and have debated whether this observation is primarily due to intrinsic size differences or to projection effects (Larsen \& Brodie 2003; Jord{\'a}n 2004). The former theory predicts a similar difference in sizes between the two subpopulations at all galactocentric distances; in the latter theory the size difference should disappear at larger projected distances. 

To create a sample of $r_h$ estimates that is as clean as possible, we imposed several criteria. First, GCs with $z<21$ were excluded, since the brighter GCs are systematically larger (see \S 3.3). Next, we excluded objects with $z>22.5$, to reduce contamination that might preferentially affect the statistics in the outermost bins. We also only used sizes from our new data, due to the small systematic difference with Jord{\'a}n \etal~(2009) as documented above (and, in general, care should be taken when combining $r_h$ estimates from different studies). This restricts the estimates to projected radii $\ga 5$ kpc.

In Figure 7 we plot the half-light radii of GCs which pass these selection criteria as a function of projected galactocentric distance. Overplotted are the modes of density estimates using a fixed Gaussian kernel of 0.25 pc, split into blue and red GC subpopulations at $g-z = 1.18$. This plot shows two interesting features. First, the metal-poor GCs are larger than the metal-rich GCs at all galactocentric distances, and the difference in $r_h$ ($\sim 0.4$ pc) does not appear to vary with distance over the range studied. This observation strongly disfavors the ``projection" hypothesis and suggests that the half-light radii of metal-poor GCs are intrinsically larger than metal-rich GCs. For reference, we have also plotted the blue and red $r_h$ modes for GCs from Jord{\'a}n \etal~(2009) within $1\arcmin$.

\begin{figure}
	\epsscale{1.2}
	\plotone{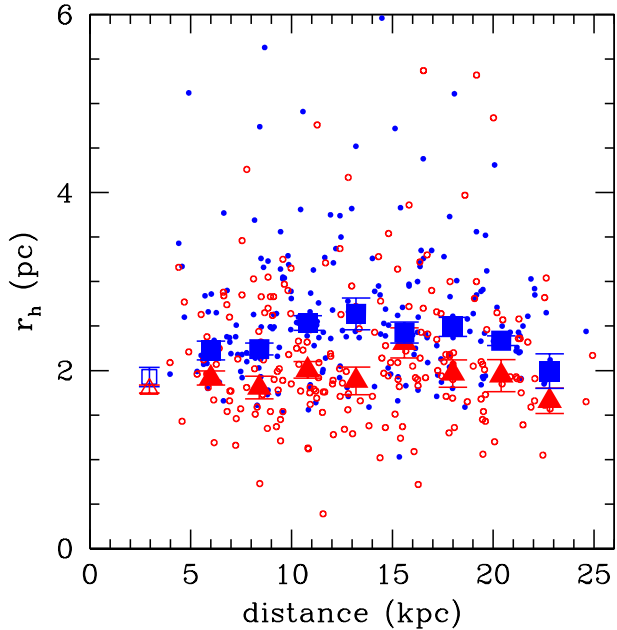}
\figcaption[f5.eps]{\label{fig:fig_7}
$r_h$ vs.~projected galactocentric distance ($R_{\rm G}$) for GCs in NGC 4649, using the selection criteria described in the text. Small blue filled circles are individual blue GCs and open circles are red GCs. The overplotted blue squares and red triangles are the modes of the metal-poor and metal-rich subpopulations. The blue GCs appear larger at all distances, and show a mild increase in $r_h$ with distance in the inner bins. At greater distances, both subpopulations show flat or declining $r_h$. The single open blue square and red triangle represent GCs from Jord{\'a}n \etal~(2009) within $1\arcmin$,
using the same selection criteria, and including the systematic offset in $r_h$ stated in \S2.3.}
\end{figure}

The other interesting feature of this plot is the run of $r_h$ with galactocentric distance. For the metal-rich GCs there is evidently no relationship over the range 5--25 kpc. For the inner ($\la 15$ kpc) metal-poor GCs there is marginal evidence for a correlation of the form: $r_h \propto R_{\rm G}^{0.14\pm0.06}$, where $R_{\rm G}$ is the projected galactocentric distance. This slope agrees with that from previous observations of other galaxies (Spitler \etal~2006; G{\'o}mez \& Woodley(2007); Harris 2009; Harris \etal~2010), though at lower significance. However, at distances $\ga 15$ kpc, the typical $r_h$ is flat or declining.

The weak inner correlation between $r_h$ and $R_{\rm G}$ and flat outer trend are consistent with another recent wide-field $HST$/ACS survey of GCs in the giant elliptical NGC 1399 (Paolillo \etal~2011). Thus there is mounting evidence that, at least in some massive galaxies, the $r_h$--$R_{\rm G}$ relation is confined to the inner regions and does not exist at all galactocentric radii. This suggests that the sizes of GCs are not generically set by tidal limitation (see also Brodie \etal~2011; Webb \etal~2012). We note that there may yet be galaxies for which the tidal limitation scenario is relevant: for example, Blom \etal~(2012) report a steep relation $r_h \propto R_{\rm G}^{0.49}$ for GCs in the elliptical NGC 4365, which is similar to the relation in the Milky Way (e.g., van den Bergh \etal~1991).

\subsection{Ultra-Compact Dwarf Galaxies}

Figure 8 shows the $r_h$ estimates of GCs in NGC 4649 as a function of $M_z$. The sizes are plotted logarithmically as the range is large. We see the correlation between $r_h$ and cluster luminosity for the most massive GCs that has been observed in many galaxies (e.g., Ha{\c s}egan \etal~2005; Evstigneeva \etal~2008; Harris 2009; Harris \etal~2010). Many of these papers have interpreted this trend as a continuous size--luminosity relationship extending from GCs to much larger ultra-compact dwarfs, and have defined the latter class solely by luminosity. Brodie \etal~(2011) argue instead that, rather than a continuous sequence of compact massive objects, there are two essentially parallel sequences separating GCs with $r_h \la 10$ pc and ultra-compact dwarfs with $r_h \ga 10$ pc over a wide range of luminosities.

The NGC 4649 data in Figure 8 are consistent with this classification: considering objects with $r_h < 10$ pc, the median size is constant at 2--3 pc for faint GCs, increasing for $M_z \la -10$, and a monotonic trend continuing to the most luminous GCs at $M_z \sim -13$. As this trend is continuous with luminosity, we see no reason to classify these most massive objects, with $r_h \sim 3$--5 pc, as anything other than very luminous star clusters. Such GCs do not appear to exist around M87---all of the objects around M87 of this luminosity are $> 10$ pc in size (Brodie \etal~2011; Strader \etal~2011)---although there are isolated examples in the literature, typically in galaxy clusters (Mieske \etal~2007; Misgeld \& Hilker 2011; Norris \& Kannappan 2011; Chiboucas \etal~2011). 

There is a parallel sequence of nearly 20 objects with $r_h \sim 10$--20 pc from $M_z \sim -8.5$ to --11. The faintest of these objects bridge the gap between the (poorly defined) faint end of the ultra-compact dwarf and the ``extended clusters" discovered in M31 and other Local Group galaxies (e.g., Huxor \etal~2011; Hwang \etal~2011). The brightest object in this size range, C28, has been confirmed as a member of NGC 4649 by previous spectroscopy (Lee \etal~2008) and is marked in Figure 8. Additional spectroscopy is necessary before further conclusions can be drawn;  it is likely a subset of these objects are background galaxies. For example, the candidates with outlying colors (e.g., E86, E123) have a larger chance of being background objects.

In addition to these moderately extended sources, there are three very large ($\sim 40$ pc) objects, two of which would fall in the ``classic" ultra-compact dwarf luminosity range $M_z \la -11$. All objects with $r_h \ga 10$ pc are listed as ultra-compact dwarf candidates in Table 2. These data emphasize the conclusion that luminosity criteria alone are insufficient to classify objects as GCs or ultra-compact dwarfs. Two of the three (A32 \& D68) have also been spectroscopically confirmed in the literature (Lee \etal~2008; Pierce \etal~2006) and thus can be recognized as bona fide ultra-compact dwarfs. They are specially marked in Figure 8.

Ultra-compact dwarfs have been discovered primarily at the centers of large galaxy clusters (e.g., Drinkwater \etal~2003), though examples exist in a range of environments (Norris \& Kannappan 2011). At least some ultra-compact dwarfs are thought to be the remnant nuclei of tidally threshed dwarf galaxies, and this process can occur most efficiently in high density environments. NGC 4649 sits at the center of a moderately massive group (Humphrey \etal~2006) in the larger context of Virgo. Given that only a small fraction of the halo radius of the galaxy has been surveyed, the confirmation of a significant number of the candidates would show that ultra-compact dwarfs can form efficiently in lower-mass environments.

\begin{figure}
	\epsscale{1.2}
	\plotone{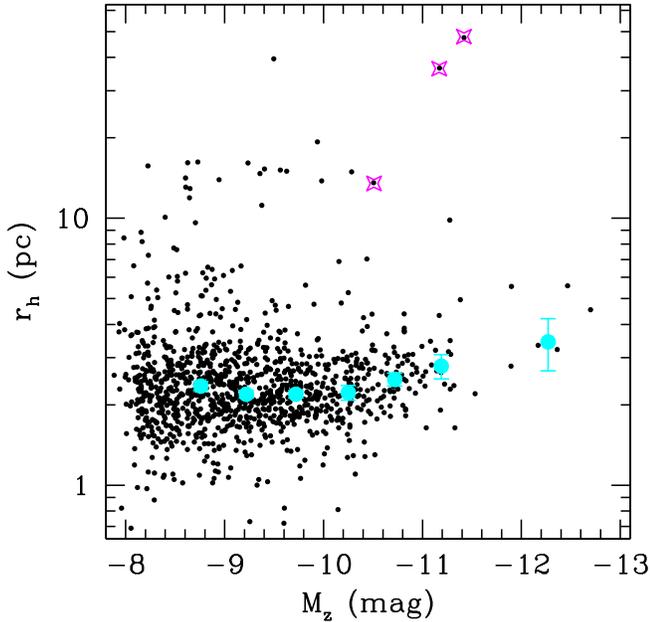}
\figcaption[n4649f6.eps]{\label{fig:fig_8}
$r_h$ vs.~$M_z$ for GCs in NGC 4649. There is a trend of increasing $r_h$ for the most luminous GCs ($M_z \la -10$) and a large population of candidate ultra-compact dwarf galaxies with $r_h > 10$ pc, extending down to surprisingly faint magnitudes. Three spectroscopically confirmed ultra-compact dwarfs are marked with magenta stars. The large cyan circles are median $r_h$ of objects with $r_h < 10$ pc in 0.5 mag bins, from $M_z =-8.5$ to $-11.5$, with uncertainties by bootstrapping. To compensate for small numbers, the final bin is contains all objects from $M_z = -11.5$ to $-13$.}
\end{figure}

\section{Summary}

We have presented photometry and half-light radii for GCs in an $HST$ mosaic of the third most massive elliptical in the Virgo cluster, NGC 4649. The improved radial coverage compared to most $HST$ studies of nearby galaxies reveals several interesting features, including significant color substructure in the metal-rich GC subpopulation at all distances and a flat or declining $r_h$ gradient beyond $\sim 15$ kpc. The metal-poor GCs are larger than the metal-rich GCs at all radii, implying that the size difference between the two subpopulations is intrinsic rather than due to projection effects. We have also identified a large population ($\ga 20$) of candidate ultra-compact dwarf galaxies that may extend to unusually faint luminosities. It is clear that the radial properties of GC systems in massive galaxies deserve further study with the high precision of $HST$.

\acknowledgments

We thank Aaron Romanowsky for helpful comments. This work has been improved by the comments of an anonymous referee. Based on observations made with the NASA/ESA \emph{Hubble Space Telescope}. This work was supported by HST Grant GO-12369.01-A (P.I~Fabbiano), Chandra Grant GO1-12110X (P.I.~Fabbiano), NASA Contract NAS8-39073 (CXC), and NSF grant AST-0808099 (P.~I.~Brodie). T.~F.~acknowledges support from the CfA and the ITC prize fellowship programs.

\LongTables

\begin{deluxetable}{lccccccccccccccc}
\tablewidth{0pt}
\tablecaption{Globular Cluster Candidates in NGC 4649
        \label{tab:phot}}
\tablehead{ID & R.A. (J2000) & Dec. (J2000) & $g$ & err & $z$ & err & $g-z$ & err & $r_{h,g}$ & err & $r_{h,z}$ & err & $r_{h,avg}$ & err & VCS?\tablenotemark{a} \\
                          & (deg) & (deg) & (mag) & (mag) & (mag) & (mag) & (mag) & (mag) & (pc) & (pc) & (pc) & (pc) & (pc) & (pc) & }

\startdata
A3 & 190.9508006 & 11.5432966 & 21.004 & 0.011 & 19.555 & 0.011 & 1.449 & 0.016 & 2.18 & 0.16 & 2.23 & 0.16 & 2.20 & 0.11 & \nodata \\
... \\
\enddata
\tablenotetext{a}{Whether the object is present in the GC catalog of Jord{\'a}n \etal~(2009).}
\end{deluxetable}

\begin{deluxetable}{lcccc}
\tablewidth{0pt}
\tablecaption{Ultra-Compact Dwarf Candidates in NGC 4649
        \label{tab:ucd}}
\tablehead{ID & $M_z$ & $g-z$ & $r_h$ & Dist.\tablenotemark{a} \\
                          & (mag)   & (mag) & (pc) & (kpc) }

\startdata

D68 & $-11.42$ & 0.99 & 47.4 & 13.1 \\
A32 & $-11.17$ & 0.92 & 36.5 & 6.2 \\
C28 & $-10.51$ & 0.93 & 13.6 & 11.5  \\
\hline
J67 & $-11.28$ & 0.88 & 9.8 & 1.8 \\
C42 & $-10.28$ & 0.92 & 14.9 & 15.4 \\
A51 & $-9.98$ & 1.66 & 13.8 & 12.9 \\
A78 & $-9.94$ & 1.56 & 19.3 & 14.2 \\
C84 & $-9.63$ & 1.59 & 15.0 & 8.8 \\
A98 & $-9.56$ & 1.26 & 15.1 & 18.7 \\
A155 & $-9.50$ & 1.03 & 39.5 & 6.1 \\
E86 & $-9.40$ & 0.61 & 15.3 & 24.9 \\
B139 & $-9.38$ & 0.95 & 11.2 & 12.6 \\
A122 & $-9.36$ & 1.60 & 14.7 & 9.0 \\
E91 & $-9.24$ & 1.69 & 16.1 & 9.3 \\
E123 & $-8.94$ & 1.85 & 13.9 & 13.0 \\
A209 & $-8.73$ & 1.01 & 16.2 & 19.5 \\
A205 & $-8.65$ & 1.23 & 12.9 & 22.5 \\
A701 & $-8.65$ & 1.73 & 11.9 & 6.2 \\
A225 & $-8.63$ & 1.04 & 16.1 & 6.9 \\
D243 & $-8.61$ & 1.60 & 13.1 & 10.2 \\
A221 & $-8.60$ & 1.33 & 14.1 & 18.5 \\
E161 & $-8.40$ & 1.00 & 10.1 & 9.7 \\
D312 & $-8.22$ & 0.83 & 15.7 & 18.7 \\

\enddata

\tablenotetext{a}{Projected galactocentric distance.}
\tablecomments{Spectroscopically confirmed objects are listed first.}

\end{deluxetable}


\begin{thebibliography}{}

\bibitem[Abazajian et al.(2009)]{2009ApJS..182..543A} Abazajian, K.~N., Adelman-McCarthy, J.~K., Ag{\"u}eros, M.~A., et al.\ 2009, \apjs, 182, 543 
\bibitem[Anderson \& Bedin(2010)]{2010PASP..122.1035A} Anderson, J., \& Bedin, L.~R.\ 2010, \pasp, 122, 1035 
\bibitem[Bertin \& Arnouts(1996)]{1996A&AS..117..393B} Bertin, E., \& Arnouts, S.\ 1996, \aaps, 117, 393 
\bibitem[Blakeslee et al.(2009)]{2009ApJ...694..556B} Blakeslee, J.~P., Jord{\'a}n, A., Mei, S., et al.\ 2009, \apj, 694, 556 
\bibitem[Blom et al.(2012)]{2012MNRAS.420...37B} Blom, C., Spitler, L.~R., \& Forbes, D.~A.\ 2012, \mnras, 420, 37 
\bibitem[Brassington et al.(2010)]{2010ApJ...725.1805B} Brassington, N.~J., Fabbiano, G., Blake, S., et al.\ 2010, \apj, 725, 1805 
\bibitem[Brodie et al.(2011)]{2011AJ....142..199B} Brodie, J.~P., Romanowsky, A.~J., Strader, J., \& Forbes, D.~A.\ 2011, \aj, 142, 199 
\bibitem[Brodie \& Strader(2006)]{2006ARAA} Brodie, J.~P., \& Strader, J.\ 2006, \araa, 44, 193
\bibitem[Chiboucas et al.(2011)]{2011ApJ...737...86C} Chiboucas, K., Tully, R.~B., Marzke, R.~O., et al.\ 2011, \apj, 737, 86 
\bibitem[C{\^o}t{\'e} et al.(2004)]{2004ApJS..153..223C} C{\^o}t{\'e}, P., Blakeslee, J.~P., Ferrarese, L., et al.\ 2004, \apjs, 153, 223 
\bibitem[de Vaucouleurs et al.(1991)]{1991rc3..book.....D} de Vaucouleurs, G., de Vaucouleurs, A., Corwin, H.~G., Jr., et al.\ 1991, Third Reference Catalogue of Bright Galaxies.
\bibitem[Drinkwater et al.(2003)]{2003Natur.423..519D} Drinkwater, M.~J., Gregg, M.~D., Hilker, M., et al.\ 2003, \nat, 423, 519 
\bibitem[Evstigneeva et al.(2008)]{2008AJ....136..461E} Evstigneeva, E.~A., Drinkwater, M.~J., Peng, C.~Y., et al.\ 2008, \aj, 136, 461 
\bibitem[Forbes et al.(2011)]{2011MNRAS.413.2943F} Forbes, D.~A., Spitler, L.~R., Strader, J., et al.\ 2011, \mnras, 413, 2943 
\bibitem[G{\'o}mez \& Woodley(2007)]{2007ApJ...670L.105G} G{\'o}mez, M., \& Woodley, K.~A.\ 2007, \apjl, 670, L105 
\bibitem[Harris et al.(2010)]{2010MNRAS.401.1965H} Harris, W.~E., Spitler, L.~R., Forbes, D.~A., \& Bailin, J.\ 2010, \mnras, 401, 1965 
\bibitem[Harris(2009)]{2009ApJ...699..254H} Harris, W.~E.\ 2009, \apj, 699, 254
\bibitem[Harris et al.(2006)]{2006ApJ...636...90H} Harris, W.~E., Whitmore, B.~C., Karakla, D., et al.\ 2006, \apj, 636, 90
\bibitem[Harris(2001)]{2001stcl.conf..223H} Harris, W.~E.\ 2001, Saas-Fee Advanced Course 28: Star Clusters, 223 
\bibitem[Ha{\c s}egan et al.(2005)]{2005ApJ...627..203H} Ha{\c s}egan, M., Jord{\'a}n, A., C{\^o}t{\'e}, P., et al.\ 2005, \apj, 627, 203 
\bibitem[Humphrey et al.(2006)]{2006ApJ...646..899H} Humphrey, P.~J., Buote, D.~A., Gastaldello, F., et al.\ 2006, \apj, 646, 899 
\bibitem[Huxor et al.(2011)]{2011MNRAS.414..770H} Huxor, A.~P., Ferguson, A.~M.~N., Tanvir, N.~R., et al.\ 2011, \mnras, 414, 770 
\bibitem[Hwang et al.(2011)]{2011ApJ...738...58H} Hwang, N., Lee, M.~G., Lee, J.~C., et al.\ 2011, \apj, 738, 58 
\bibitem[Irwin et al.(2010)]{2010ApJ...712L...1I} Irwin, J.~A., Brink, T.~G., Bregman, J.~N., \& Roberts, T.~P.\ 2010, \apjl, 712, L1 
\bibitem[Jord{\'a}n et al.(2009)]{2009ApJS..180...54J} Jord{\'a}n, A., Peng, E.~W., Blakeslee, J.~P., et al.\ 2009, \apjs, 180, 54 
\bibitem[Jord{\'a}n et al.(2007)]{2007ApJS..171..101J} Jord{\'a}n, A., McLaughlin, D.~E., C{\^o}t{\'e}, P., et al.\ 2007, \apjs, 171, 101 
\bibitem[Jord{\'a}n et al.(2005)]{2005ApJ...634.1002J} Jord{\'a}n, A., C{\^o}t{\'e}, P., Blakeslee, J.~P., et al.\ 2005, \apj, 634, 1002 
\bibitem[Jord{\'a}n(2004)]{2004ApJ...613L.117J} Jord{\'a}n, A.\ 2004, \apjl, 613, L117 
\bibitem[Kundu et al.(2002)]{2002ApJ...574L...5K} Kundu, A., Maccarone, T.~J., \& Zepf, S.~E.\ 2002, \apjl, 574, L5 
\bibitem[Kundu \& Whitmore(1998)]{1998AJ....116.2841K} Kundu, A., \& Whitmore, B.~C.\ 1998, \aj, 116, 2841 
\bibitem[Larsen \& Brodie(2003)]{2003ApJ...593..340L} Larsen, S.~S., \& Brodie, J.~P.\ 2003, \apj, 593, 340 
\bibitem[Larsen et al.(2001)]{2001AJ....121.2974L} Larsen, S.~S., Brodie, J.~P., Huchra, J.~P., Forbes, D.~A., \& Grillmair, C.~J.\ 2001, \aj, 121, 2974 
\bibitem[Larsen(1999)]{1999A&AS..139..393L} Larsen, S.~S.\ 1999, \aaps, 139, 393 
\bibitem[Lee et al.(2008)]{2008ApJ...674..857L} Lee, M.~G., Hwang, H.~S., Park, H.~S., et al.\ 2008, \apj, 674, 857 
\bibitem[Luo et al.(2012)]{Luo12} Luo, B., Fabbiano, G., et al.\ 2012, in preparation
\bibitem[Maccarone et al.(2007)]{2007Natur.445..183M} Maccarone, T.~J., Kundu, A., Zepf, S.~E., \& Rhode, K.~L.\ 2007, \nat, 445, 183 
\bibitem[Misgeld \& Hilker(2011)]{2011MNRAS.414.3699M} Misgeld, I., \& Hilker, M.\ 2011, \mnras, 414, 3699 
\bibitem[Mieske et al.(2010)]{2010ApJ...710.1672M} Mieske, S., Jord{\'a}n, A., C{\^o}t{\'e}, P., et al.\ 2010, \apj, 710, 1672 
\bibitem[Mieske et al.(2007)]{2007A&A...472..111M} Mieske, S., Hilker, M., Jord{\'a}n, A., Infante, L., \& Kissler-Patig, M.\ 2007, \aap, 472, 111 
\bibitem[Mieske et al.(2006)]{2006ApJ...653..193M} Mieske, S., Jord{\'a}n, A., C{\^o}t{\'e}, P., et al.\ 2006, \apj, 653, 193 
\bibitem[Norris \& Kannappan(2011)]{2011MNRAS.414..739N} Norris, M.~A., \& Kannappan, S.~J.\ 2011, \mnras, 414, 739 
\bibitem[Paolillo et al.(2011)]{2011ApJ...736...90P} Paolillo, M., Puzia, T.~H., Goudfrooij, P., et al.\ 2011, \apj, 736, 90
\bibitem[Peng et al.(2006)]{2006ApJ...639...95P} Peng, E.~W., Jord{\'a}n, A., C{\^o}t{\'e}, P., et al.\ 2006, \apj, 639, 95 
\bibitem[Pierce et al.(2006)]{2006MNRAS.368..325P} Pierce, M., Bridges, T., Forbes, D.~A., et al.\ 2006, \mnras, 368, 325 
\bibitem[Pooley et al.(2003)]{2003ApJ...591L.131P} Pooley, D., Lewin, W.~H.~G., Anderson, S.~F., et al.\ 2003, \apjl, 591, L131 
\bibitem[Schlegel et al.(1998)]{1998ApJ...500..525S} Schlegel, D.~J., Finkbeiner, D.~P., \& Davis, M.\ 1998, \apj, 500, 525 
\bibitem[Sirianni et al.(2005)]{2005PASP..117.1049S} Sirianni, M., Jee, M.~J., Ben{\'{\i}}tez, N., et al.\ 2005, \pasp, 117, 1049 
\bibitem[Spitler et al.(2006)]{2006AJ....132.1593S} Spitler, L.~R., Larsen, S.~S., Strader, J., Brodie, J.~P., Forbes, D.~A., \& Beasley, M.~A.\ 2006, \aj, 132, 1593 
\bibitem[Strader et al.(2011)]{2011ApJS..197...33S} Strader, J., Romanowsky, A.~J., Brodie, J.~P., et al.\ 2011, \apjs, 197, 33 
\bibitem[Strader et al.(2006)]{2006AJ....132.2333S} Strader, J., Brodie, J.~P., Spitler, L., \& Beasley, M.~A.\ 2006, \aj, 132, 2333 
\bibitem[van den Bergh et al.(1991)]{1991ApJ...375..594V} van den Bergh, S., Morbey, C., \& Pazder, J.\ 1991, \apj, 375, 594 
\bibitem[van Dokkum(2001)]{2001PASP..113.1420V} van Dokkum, P.~G.\ 2001, \pasp, 113, 1420 
\bibitem[Webb et al.(2012)]{2012ApJ...746...93W} Webb, J.~J., Sills, A., \& Harris, W.~E.\ 2012, \apj, 746, 93 

\end{thebibliography}
\end{document}